\documentclass[amsmath,amsfonts,twocolumn,superscriptaddress,prx,longbibliography]{revtex4-1}
\usepackage{graphicx}
\usepackage{epsfig}
\usepackage{bm}
\usepackage{dcolumn}
\usepackage{amsmath}
\usepackage{amssymb}
\usepackage{gensymb}
\usepackage{color}
\usepackage{textgreek}

\let\a=\alpha \let\b=\beta \let\g=\gamma \let\d=\partial  \let\l=\lambda 
\let\k=\kappa \let\s=\sigma \let\t=\tau \def\vep{{\varepsilon}}

\let\D=\Delta \let\G=\Gamma

\newcommand{\Tr}{\mathop{\rm Tr}\nolimits}

\def\ep{{\epsilon}}

\begin{document}

\title{Non-Abelian monopoles in the multiterminal Josephson effect}

\author{Hong-Yi Xie}
\affiliation{Division of Quantum State of Matter, Beijing Academy of Quantum Information Sciences, Beijing 100193, China}
 
\author{Jaglul Hasan}
\affiliation{Department of Physics, University of Wisconsin--Madison, Madison, Wisconsin 53706, USA} 
 
\author{Alex Levchenko}
\affiliation{Department of Physics, University of Wisconsin--Madison, Madison, Wisconsin 53706, USA} 

\begin{abstract}
In this work we present a detailed theoretical analysis for the spectral properties of Andreev bound states in the multiterminal Josephson junctions by employing a symmetry-constrained scattering matrix approach. We find that in the synthetic multidimensional space of superconducting phases, crossings of Andreev bands may support non-Abelian SU(2) monopoles with a topological charge characterized by the second class Chern number. We propose that these topological defects can be detected via a nonlinear response measurement of the current autocorrelations. In addition, multiterminal Josephson junction devices can be tested as a hardware platform for realizing holonomic quantum computation.
\end{abstract}

\date{June 6, 2022}

\maketitle

\section{Introduction} 

Topological defects, such as domain walls, vortices, monopoles, skyrmions, etc., play a special role in physics and lead to a number of fascinating phenomena with nonperturbative effects. In particular, divergent configurations of the monopole field strength generate quantized flux through any manifold enclosing singular point, which is stable to deformations. The corresponding invariant -- topological charge --  can be classified by the Chern numbers \cite{Chern}. Topologies associated with the first Chern class are abundant in physics realizations, most notably found in the context of quantized Hall conductance \cite{TKNN}. However, topologies of the second Chern class are more elusive as they reside in higher-dimensional spaces. The focus of this work is on the Yang's monopole \cite{Yang} that was originally introduced in the context of non-Abelian gauge fields in five-dimensional flat space and reemerged in the condensed matter theory construction of the four-dimensional (4D) quantum Hall effect \cite{Zhang-Hu}. At present, practical realizations of the Yang monopole were discussed from the perspectives of the spin Hall effect in hole-doped semiconductors \cite{Murakami}, states in quasicrystals \cite{Kraus}, higher-spin fermionic superfluids \cite{ChernChen}, and Bose-Einstein condensates \cite{Sugawa}. The proposed measurement protocols for the second Chern number can be potentially implemented based on spin-$\frac{3}{2}$ particles in an electric quadrupole field \cite{Kolodrubetz}. 

In the recent years multiterminal Josephson junctions were proposed as a promising hardware platform for creating topological states including higher-dimensional topologies \cite{Riwar}. This idea sparked tremendous interest followed by a multitude of studies \cite{Eriksson,Xie1,MH1,Xie2,Deb,Erdmanis,MH2,Xie3,Kotetes,Pekker,Kornich,Gavensky,Marra,Repin1,Sakurai,Regis1,Klees,MH3,Fatemi,Chirolli,Peyruchat,Repin2,Septembre,ChenNazarov,Regis2} that cover a broad spectrum of device designs, transport properties, material components, as well as extensions to the higher-order topologies \cite{Weisbrich,Belzig}. In part the interest is also motivated by the perspective applications to quantum computation and realization of the holonomic gates that require adiabatic control of the system driven across the parameter space with a non-Abelian Berry connection \cite{Pachos}. In the multiterminal superconducting circuits monopoles correspond to Weyl or Majorana zero-energy crossings of Andreev bound states localized in the subgap region. Depending on a particular model realization some of these crossings may be lifted in energy. A crucial point is that these topologies are enabled by the design of the device. The occurrence of a topological crossing can be tested numerically by generating random scattering matrices and a statistically significant fraction of scattering matrices was found to yield Weyl points \cite{Riwar}. This analysis suggests that topological spectral features remain robust at the mesoscopic level. They can be further enriched if the material constituents forming the junction possess intrinsic topological properties. 

The key predicted signatures of the topological regime in transport properties are the quantized nonlocal conductance \cite{Eriksson,Xie2,MH3}, in fundamental units of $(4e^2/h)\mathcal{C}_1$, as well as adiabatically quantized charge pumping \cite{Pekola,ChenNazarov,Leone}, in units of $2e\mathcal{C}_1$ per winding around the monopole. Here, $e$ is the electric charge, $h$ is the Planck constant, and $\mathcal{C}_1$ represents the first class Chern number of the underlying Berry curvature flux of nontrivial Andreev bands. The experimental efforts in creating multiply connected superconducting circuits include Andreev interferometers \cite{Strambini}, proximitized graphene \cite{Draelos,Arnault}, hybrid superconducting-semiconductor epitaxial heterostructure, and nanowires \cite{Cohen,Pankratova,Graziano,Pribiag}.   

In this work, we explore the idea of a multiterminal Josephson effect as a practical platform for realizing topological artificial matter and construct an analytically solvable model that captures the properties of the Yang's monopole. We derive spectral properties of the prototypical device and demonstrate the non-Abelian character of the obtained band structure. For this purpose, in Sec. \ref{sec:S-matrix} we use the symmetry-constrained scattering matrix approach with the specific parametrization of unitary matrices. It enables us to take advantage of the properties of palindromic polynomials to resolve analytically the eigenvalue problem. 
This way we calculate the Andreev spectrum of five- and six-terminal Josephson junction devices. To establish connection to the experimentally detectable responses, possible transport signatures of the multiterminal Josephson junctions in the topological non-Abelian phase are briefly discussed in Sec. \ref{sec:transport}. 

\section{$S$-matrix formalism}\label{sec:S-matrix}

We consider a mesoscopic $N$-terminal Josephson junction where all superconducting leads share the same normal scattering region. We assume each lead to be a conventional $s$-wave superconductor (SC) with a rigid energy gap $\D e^{i \theta_\a}$ and the whole device to have a nonuniform set of distributed phases $\theta_\a$, with $\a = 0,\cdots,N-1$ labeling the corresponding terminal. Thus each phase plays an effective role of an artificial dimension. For simplicity, we take the energy-independent scattering matrix $\hat{s}$ of a normal bridge with a single channel per spin per lead. This approximation describes well short junctions where the typical size $L$ of the contact is small as compared to the superconducting coherent length $L\ll\xi$. In this setting, we introduce the normal-state propagating wave basis $\psi_{\pm}^{\a} = e^{\pm i k_{\a} x_\a}/\sqrt{2\pi\hbar v_\a}$, where $k_\a$ is the Fermi momentum and $v_\a$ is the Fermi velocity, and $\pm$ denotes the incident-to-lead (reflected-from-lead) component. The four-spinor quasiparticle wave function of energy $E$ at the $\a^{\text{th}}$ lead $\Psi^\a(E)$ is a linear combination of propagating waves 
\begin{equation}
\Psi^\a(E) \equiv 
\begin{pmatrix} \Psi_{\mathrm{e}\downarrow}^\a \\ \Psi_{\mathrm{h}\uparrow}^\a \\ \Psi_{\mathrm{e}\uparrow}^\a \\ \Psi_{\mathrm{h}\downarrow}^\a \end{pmatrix} 
= \begin{pmatrix} 
A_{\mathrm{e}\downarrow}^{\a+} \psi_{+}^{\a} + A_{\mathrm{e}\downarrow}^{\a-}  \psi_{-}^{\a}  \\  
A_{\mathrm{h}\uparrow}^{\a+}   \psi_{-}^{\a} + A_{\mathrm{h}\uparrow}^{\a-}    \psi_{+}^{\a}  \\
A_{\mathrm{e}\uparrow}^{\a+}   \psi_{+}^{\a} + A_{\mathrm{e}\uparrow}^{\a-}    \psi_{-}^{\a}  \\  
A_{\mathrm{h}\downarrow}^{\a+}  \psi_{-}^{\a}+ A_{\mathrm{h}\downarrow}^{\a-}  \psi_{+}^{\a}
 \end{pmatrix},
\end{equation}
graded by pseudospin $\otimes$ particle-hole space, where the coefficients $\{A_{\t,\s}^{\a,\pm}\}$ are the scattering amplitudes. In this basis representation \cite{beenakkerRMP}, the charge and spin density operator respectively take the expressions 
\begin{equation}  
\rho =  -\frac{e}{2} \s_0 \otimes \t_3, \quad 
\boldsymbol{\Sigma} = \frac{\hbar}{4} \begin{pmatrix} \s_1 \otimes \t_3, & -\s_2 \otimes \t_3, & -\s_3\otimes\t_0  \end{pmatrix},
\end{equation}
where $\s_{1,2,3}$ and $\t_{1,2,3}$ are sets of Pauli matrices operational in pseudo-spin and particle-hole spaces, respectively. 
In general, the Andreev bound state (ABS) amplitudes $\mathbf{A}^\pm(E)$ with $|E|<\Delta$ are the eigenstates of the scattering matrix belonging to the eigenvalue ``1'' \cite{BeenakkerPRL},
\begin{equation} \label{abs-0}
\big[ \mathbb{I}_{4N} - U(E) \big] \mathbf{A}^{+}(E) = 0, \quad  U(E) \equiv S R(E),  
\end{equation}
and $\mathbf{A}^{-}(E) = R(E) \mathbf{A}^{+}(E)$, where the scattering matrix $S$ of the normal region and the Andreev reflection boundary matrix $R(E)$ respectively read
\begin{align}
& S = \s_0 \otimes \begin{pmatrix} s & 0 \\ 0 & s^\ast \end{pmatrix}, \quad
  R(E) = \begin{pmatrix} r(E) & 0 \\ 0 & \tau_3 r(E) \tau_3 \end{pmatrix}. \label{SR}
\end{align}
In Eq.~\eqref{SR}, $s \in \mathrm{SU}(N)$, and $r(E)$ is a unitary matrix of which the particle-hole space blocks are diagonal $r_{\t\t'}(E) = \mathrm{diag}\{r_{\t\t'}^0(E),\ldots,r_{\t\t'}^{N-1}(E)\}$, where $r_{\t\t'}^\a(E)$ describes the $\t' \to \t$ reflection amplitude at lead $\a$. The particle-hole symmetry is represented by
\begin{equation} \label{phs}
\mathcal{P}^{-1} U(-E) \mathcal{P} = U(E), \quad \mathcal{P} \equiv i \s_0 \otimes \t_2 \mathcal{K},
\end{equation} 
where $\mathcal{K}$ is the complex conjugation and $\mathcal{P}^2=-1$. The spin SU(2) rotation symmetry is represented by 
\begin{equation} \label{srs}
\mathcal{S}^{-1} U(E) \mathcal{S} = U(E), \quad \mathcal{S} \equiv e^{i \boldsymbol{\Sigma} \cdot \boldsymbol{\eta}/\hbar}. 
\end{equation} 

Within the Andreev approximation, namely neglecting the normal reflections at NS interfaces, the Andreev bound states can be obtained by solving the eigenvalue problem of a unitary matrix. The scattering matrix $U(E)$ in Eq.~\eqref{abs-0} can be simplified to $U(\ep) = \g(\ep) \s_3 \otimes  Q$, where
\begin{align}
& \g(\ep) = e^{-i \arccos \ep}, \quad Q = \begin{pmatrix} 0 & s e^{i \hat{\theta}} \\ s^\ast e^{-i\hat{\theta}} & 0 \end{pmatrix}, \label{Q-mat}
\end{align} 
with $\ep=E/\D \in [-1,1]$ and $\hat{\theta} = \mathrm{diag}\{\theta_0,\ldots,\theta_{N-1}\}$. The ABS equation \eqref{abs-0} reduces to $\left[ \mathbb{I}_{2N} \mp \g(\ep) Q \right] \mathbf{A}_{\pm} = 0$, where $\mathbf{A}_{+}$ and $\mathbf{A}_{-}$ correspond to the eigenvalues of $Q$ with the phases in the interval $[0,\pi]$ and $[-\pi,0]$, respectively (see analogous details in Ref. \cite{Akhmerov}). 

We introduce an effective Hamiltonian of the Andreev bound states by $H(\boldsymbol{\theta}) \equiv (Q + Q^{\dagger})/2$ and thus obtain
\begin{align} 
& H(\boldsymbol{\theta}) = \begin{pmatrix} 0 & D(\boldsymbol{\theta}) \\ D^\dagger(\boldsymbol{\theta}) & 0 \end{pmatrix}, \quad 
  D(\boldsymbol{\theta}) = \G_0 + e^{i \boldsymbol{\theta}} \cdot \boldsymbol{\G}, 
\label{abs-ham}
\end{align} 
where the $\{\G_\a\}$ matrices are symmetric and the elements read $\G_{\a}^{\mu \nu} = (s_{\mu \nu} \delta_{\nu \a}+ s_{\nu \mu} \delta_{\mu\a})/2$ with $\delta_{\mu\nu}$ being the Kronecker delta function, and we have set the $N$th phase to $\theta_0 = 0$ owning to the global gauge invariance and denoted $\boldsymbol{\theta} \equiv (\theta_1,\ldots,\theta_{N-1})$. We note that in addition to the particle-hole symmetry Eq.~\eqref{phs}, the Hamiltonian satisfies the chiral symmetry, $\{\mathcal{C};H\} = 0$ with $\mathcal{C} = \t_3$, and thus a combined time-reversal symmetry (TRS), $[\mathcal{T};H] = 0$ with $\mathcal{T} \equiv \mathcal{C} \mathcal{P} = \tau_1 \mathcal{K}$ and $\mathcal{T}^2=1$. Therefore, the Hamiltonian \eqref{abs-ham} belongs the Altland-Zirnbauer class CI \cite{AZ}. The Andreev bound states are given by $H(\boldsymbol{\theta}) |\Phi_{n}^\kappa \rangle = \ep_n(\boldsymbol{\theta}) |\Phi_{n}^\kappa \rangle$, for which each band is doubly degenerate with two eigenstates $\kappa=1,2$ related by the outlined symmetries, $|\Phi_{n}^1 \rangle = \mathcal{T} |\Phi_{n}^2 \rangle$. 

\subsection{Andreev bound states spectrum} 

The Andreev spectrum is determined by the roots of the $Q$-matrix characteristic polynomial $P_N(\g) \equiv \mathrm{Det}(\mathbb{I}_{2N} - \g Q)$. For the $Q$ matrix in Eq.~\eqref{Q-mat} we obtain for the determinant 
\begin{equation} \label{q-mat}
P_N(\g) = \mathrm{Det}(\mathbb{I}_{N} - \g^2 q), \quad q \equiv s^\ast e^{-i\hat{\theta}}s e^{i \hat{\theta}}, 
\end{equation} 
that is a palindromic (antipalindromic) polynomial of $\g^2$ for $N \in$ even (odd). For the five-terminal ($N=5$) junctions, we obtain four nontrivial Andreev bands
\begin{equation} \label{ABS-5T}
\vep (\boldsymbol{\theta}) = \pm \sqrt{\frac{4 + A_5 \pm \sqrt{A_5^2 -4 B_5 +8}}{8}},
\end{equation}  
and one trivial band $\vep(\boldsymbol{\theta})=1$, where $\boldsymbol{\theta} \equiv (\theta_1,\theta_2,\theta_3,\theta_4)$, and the functions $A_5(\boldsymbol{\theta})$ and $B_5(\boldsymbol{\theta})$ are determined by the $q$ matrix in Eq.~\eqref{q-mat} as follows: 
\begin{align} 
A_5 = \Tr q-1, \quad 
B_5  = \frac{1}{2} \Tr^2{q} - \frac{1}{2} \Tr{q}^2 -\Tr{q}+1.    
\end{align}
For the six-terminal ($N=6$) junctions, we obtain three pairs of Andreev bands
\begin{align}
\vep_m (\boldsymbol{\theta}) = \pm 
\sqrt{ \frac{6+A_6 - 2 \sqrt{A_6^2-3 B_6+9} \, \cos\left( \frac{\Phi + 2 m \pi}{3} \right)}{12}},
\end{align} 
where $m \in \{0,1,2\}$, $\boldsymbol{\theta} \equiv (\theta_1,\theta_2,\theta_3,\theta_4,\theta_5)$, the $\Phi$ function reads  
\begin{equation}  \label{ABS-6T}
\Phi(\boldsymbol{\theta}) \equiv \arccos\left[ \frac{-2 A_6^3 + 9 A_6 B_6 + 27 (A_6- C_6)}{2 (A_6^2-3 B_6+9)^{3/2}} \right],
\end{equation} 
and the functions $A_6$, $B_6$, and $C_6$ are given by expressions 
\begin{align} 
A_6 = &\, \Tr q, \quad B_6 = \frac{1}{2} \Tr^2q - \frac{1}{2} \Tr q^2, \nonumber \\
C_6 = &\, \frac{1}{6} \Tr^3 q + \frac{1}{3} \Tr q^3 - \frac{1}{2} \Tr q \, \Tr q^2.
\end{align}

We choose to parametrize the $\hat{s}$-matrix according to the decomposition $\hat{s} = \hat{u} e^{i \hat{d}} \hat{u}^\dagger$, where $\hat{d}$ is real diagonal and $\hat{u}$ is unitary. For concreteness we analyze a special realization with matrix elements of the form 
\begin{equation} \label{d_alp}
d_\a=2 \mathrm{arctan}\left[ \mu + 2 t \cos\left(\frac{\phi+2\pi \a}{N}\right) \right], \,\, u_{\a\b} = \frac{ e^{i \a \b \frac{2\pi}{N}}}{\sqrt{N}}, 
\end{equation} 
where $\a,\b$ take values from 0 to $N-1$ and $d_\a \in [-\pi,\pi]$. Hence, $s_{\a\b} =  \frac{1}{N}\sum_{\nu} e^{i[2\pi \nu (\a-\b)/N + d_\nu]}$. This $\hat{s}$-matrix satisfies the $N$-polygon symmetry and the three free parameters $\mu$, $t$, and $\phi$ represent the on-site chemical potential, on-site hopping energy, and overall flux through the polygon area, respectively, analogous to the single-site multiterminal junction model introduced in Ref.~\cite{MH1}. As the free parameters varying, the Andreev band gaps can close only at the commensurate SC phases $\theta_\a^{(n)} \equiv \mathrm{mod}( \a n 2\pi/N,\, 2\pi)$ with $0 \le n \le N-1$. At these special phase points, we introduce the quantities $T_{n,p}^{{N}} \equiv \Tr q^p(\hat{\theta}^{(n)})$ with $p \in \mathbb{N}$, that determine the ABS energies in Eqs. \eqref{ABS-5T} and \eqref{ABS-6T}. The result distinguishes between even and odd numbers of terminals. 

\begin{figure}
\includegraphics[width=0.235\textwidth]{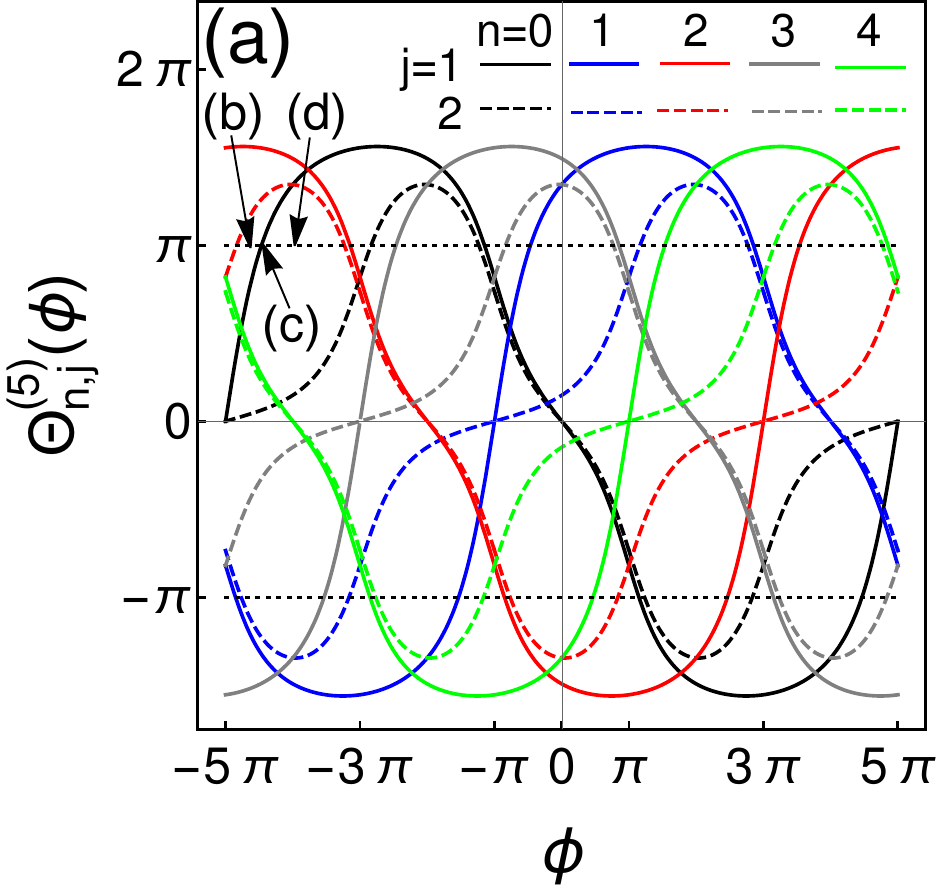} \,\,
\includegraphics[width=0.225\textwidth]{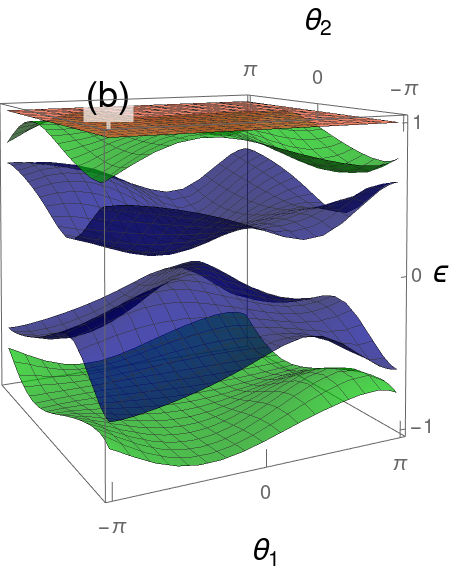} \\
\includegraphics[width=0.225\textwidth]{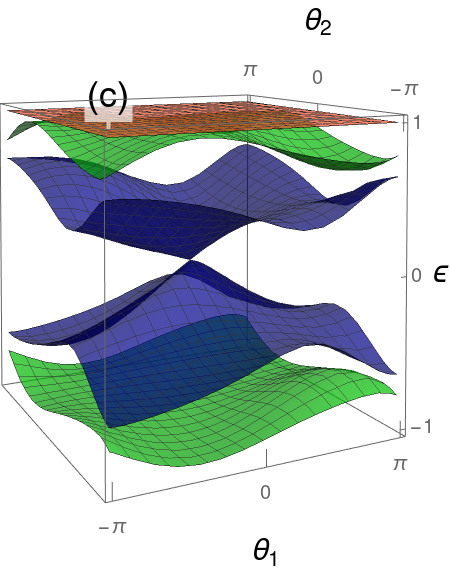} \,\,
\includegraphics[width=0.225\textwidth]{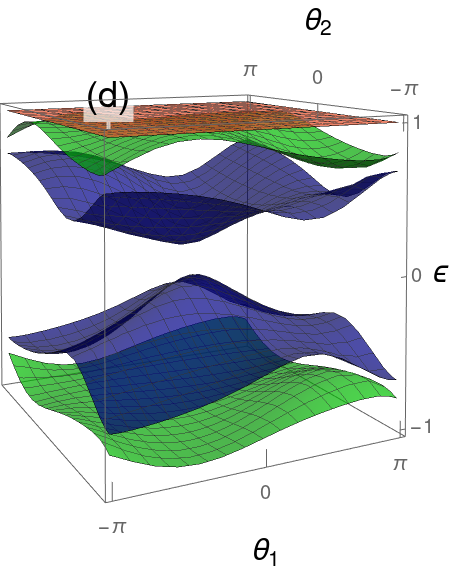}
\caption{
ABS spectrum for a model example of $N=5$ terminal Josephson junction for various $\phi$ and fixed $\mu$ and $t$. We take $\mu=1$ and $t=1.6$ [Eqs.~\eqref{ABS-5T} and \eqref{d_alp}]. There is one trivial branch at $\ep=1$. (a) Phase differences $\{\Theta_{n,j}^{(5)}\}$ in Eq.~\eqref{Theta-5} as functions of $\phi$. (b)-(d) ABS spectrum as a function of $\theta_{1,2}$ for $\theta_{3,4}=0$ for various values of $\phi$ as labeled in panel (a). (b) $\phi=-14.2$. (c) $\phi \approx -14.0301$. A Dirac node forms at $\theta_{1,2} =0$. (d) $\phi=-13.8$.
}  \label{fig1}
\end{figure}

(i) For the odd number of terminals ($N \ge 3$), we obtain
\begin{equation} \label{Theta-5}
T_{n,p}^{{N}}  = 1 + 2 \sum_{j=1}^{\mathsf{N}-1} \cos\left[ p \Theta_{n,j}^{(N)} \right],
\end{equation}
where $\mathsf{N} \equiv (N+1)/2$ and $\Theta_{n,j}^{(N)} =  d_{\langle \langle n \mathsf{N} \rangle+j \rangle}-d_{\langle \langle n \mathsf{N} \rangle-j \rangle}$ with $\langle n \rangle \equiv \mathrm{mod}(n,N)$. Specifically, for $N=5$ a zero-energy Dirac point forms at $\hat{\theta}^{(n)}$ when $\Theta_{n,j}^{(5)}=\pm\pi$. We show $\{\Theta_{n,j}^{(5)}\}$ and the Andreev spectrum as functions of $\phi$ for fixed $\mu$ and $t$ in Figs.~\ref{fig1}(a)--\ref{fig1}(d). 

\begin{figure}
\includegraphics[width=0.235\textwidth]{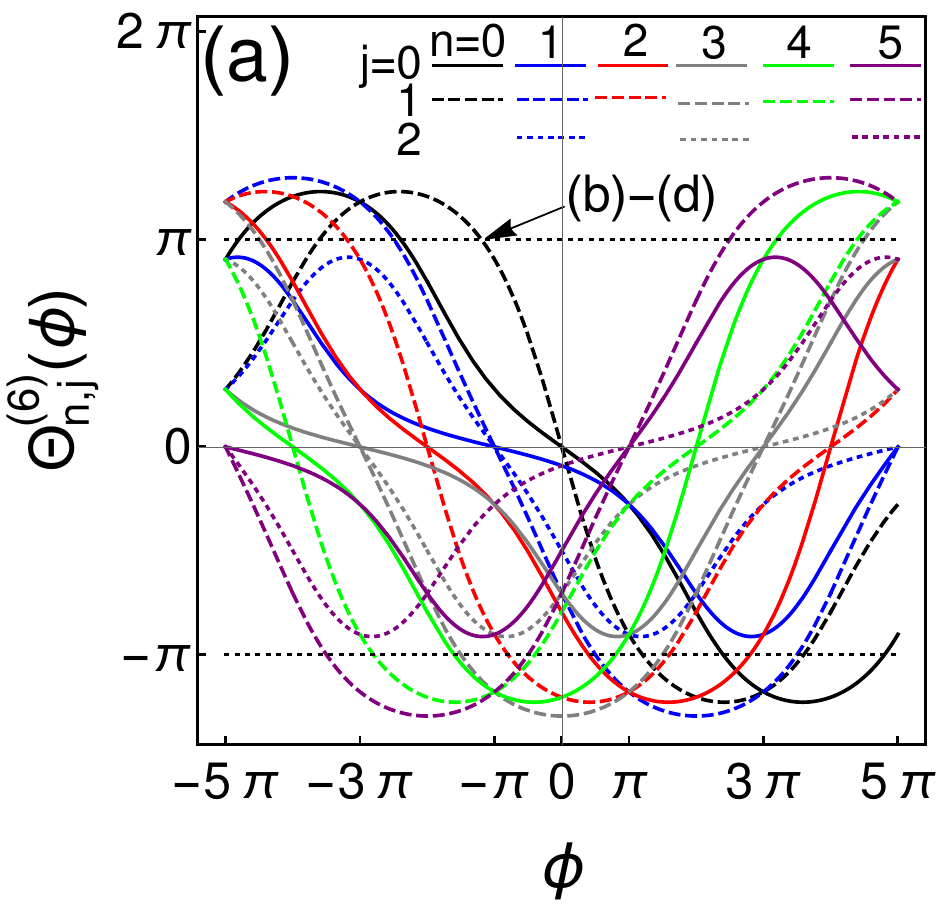} \,\,
\includegraphics[width=0.225\textwidth]{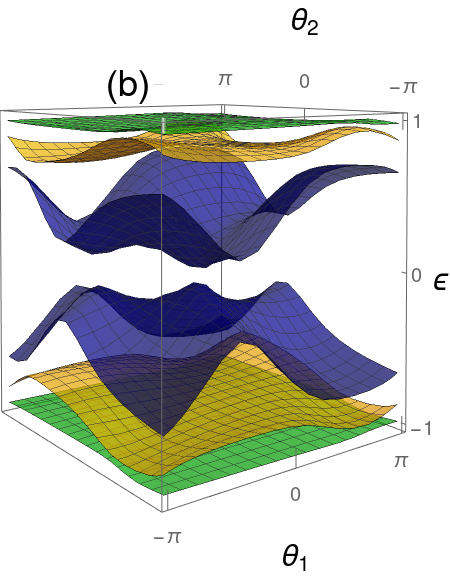} \\
\includegraphics[width=0.225\textwidth]{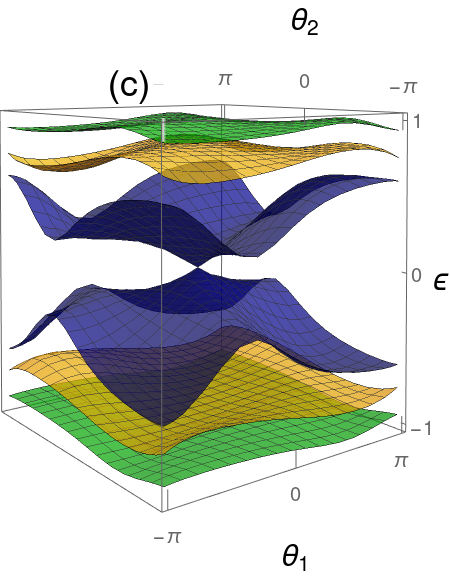} \,\,
\includegraphics[width=0.225\textwidth]{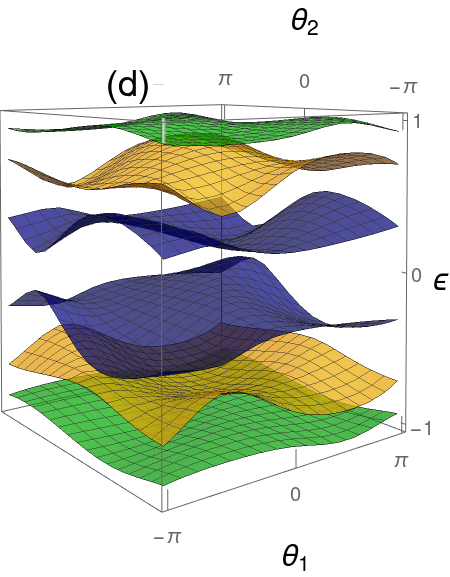}
\caption{ABS spectrum of $N=6$ terminal Josephson junctions for a fixed $s$ matrix.
(a) Phase differences $\{\Theta_{n,j}^{(6)}\}$ in Eq.~\eqref{Theta-6} as functions of $\phi$. A Yang monopole forms at $\boldsymbol{\theta}=0$ for $\mu=t=1$ and $\phi \approx -3.7699$. (b)-(d) ABS spectrum as a function of $\theta_{1,2}$ for $\theta_{4,5}=0$ for various values of $\theta_3$. (b) $\theta_3=-\pi/3$. (c) $\theta_3=0$. The monopole manifests. (d) $\theta_3=\pi/3$.
}  \label{fig2}
\end{figure}

(ii) For the even number of terminals ($N \ge 4$), the result is distinguished by the even and odd values of $n$,
\begin{align}
T_{n,p}^{{N}} = 
\begin{cases} 
2 + 2 \displaystyle \sum_{j=1}^{\mathsf{N}-1} \cos\left[ p \, \Theta_{n,j}^{(N,\mathrm{e})} \right], \quad & n \in \mathrm{even}, \\ 
2 \displaystyle \sum_{j=0}^{\mathsf{N}-1} \cos\left[ p \, \Theta_{n,j}^{(N,\mathrm{o})} \right],     \quad & n \in \mathrm{odd},
\end{cases}  \label{Theta-6}
\end{align}
where $\mathsf{N} \equiv N/2$, $\Theta_{n,j}^{(N,\mathrm{e})} = d_{\langle \frac{n}{2} +j \rangle}-d_{\langle \frac{n}{2} -j \rangle}$, and 
$\Theta_{n,j}^{(N,\mathrm{o})} = d_{\langle \frac{n+1}{2} +j \rangle}-d_{\langle \frac{n-1}{2}-j \rangle}$. 
Specifically, for $N=6$ a zero-energy Dirac point forms at $\hat{\theta}^{(n)}$ when $\Theta_{n,j}^{(6,\mathrm{e}/\mathrm{o})}=\pm\pi$. 
We depict $\{\Theta_{n,j}^{(6,\mathrm{e}/\mathrm{o})}\}$ as functions of $\phi$ for fixed $\mu$ and $t$ in Fig.~\ref{fig2}(a). For the specific values of $\mu$, $t$, and $\phi$, a monopole forms at $\boldsymbol{\theta}=0$ and we show the Andreev spectrum in Figs.~\ref{fig2}(b)--\ref{fig2}(d). We note that alternatively one could set $\phi\to0$ and vary the parameter space defined by $\mu$ and $t$ to achieve a topological regime.

\subsection{Second Chern number} 

We proceed to study the topological phases of five -and six-terminal junctions characterized by the second Chern number of the Andreev bound states $\{|\Phi_n^{\kappa} (\boldsymbol{\theta}) \rangle \}$, that are the eigenstates of the Hamiltonian \eqref{abs-ham}. 
For $N \ge 5$, we define the states on a four-dimensional torus $\theta_{1 \le \a \le 4} \in [0,2\pi]$ and consider the remaining SC phases $\theta_{\a \ge 5}$ as fixed parameters if $N >5$. For the $n$th band, we can define the U(2) Berry connection \cite{WilczekZee},
\begin{align} 
\mathcal{A}_{n,\a}^{\k\k'}  (\boldsymbol{\theta})  \equiv -i \big\langle \Phi_n^\k \big| \d_\a \big| \Phi_n^{\k'}  \big\rangle, \quad \k,\k'=1,2, \label{bcc}
\end{align} 
where $\d_\a \equiv \d_{\theta_\a}$, which can be decomposed into U(1) and SU(2) parts as $\hat{\mathcal{A}}_{n,\a} = (\mathfrak{a}_{n,\a}^0 \hat{\k}^0 + \mathfrak{a}_{n,\a}^j \hat{\k}^j)/2$ and the U(1) part vanishes $\mathfrak{a}_{n,\a}^0=0$ in the case of time-reversal symmetry. The corresponding Berry curvature is defined as 
\begin{equation} \label{bcv}
\hat{\mathcal{F}}_{n,\a\b} 
\equiv \d_\a \hat{\mathcal{A}}_{n,\b}- \d_\b \hat{\mathcal{A}}_{n,\a} + i \big[ \hat{\mathcal{A}}_{n,\a}, \hat{\mathcal{A}}_{n,\b} \big]=\frac{\mathfrak{f}_{n,\a\b}^j}{2}  \hat{\k}_j, 
\end{equation}
where $\mathfrak{f}_{n,\a\b}^j  = \d_\a \mathfrak{a}_{n,\b}^j - \d_\b \mathfrak{a}_{n,\a}^j - \varepsilon^{jkl} \mathfrak{a}_{n,\a}^k \mathfrak{a}_{n,\b}^l$. 
With these notations we can explore the analogy with the Yang-Mills gauge theory 
and introduce the second Chern number of the $n$ band that reads \cite{BPST,Witten,Yang}
\begin{equation} \label{chern2}
\mathcal{C}^{[n]}_{2} \equiv \oint_{\boldsymbol{\theta}} \!  \varepsilon^{\a\b\g\delta} \Tr \big[ \hat{\mathcal{F}}_{n,\a\b} \hat{\mathcal{F}}_{n,\g\delta} \big]
                  = \oint_{\boldsymbol{\theta}} \!  \varepsilon^{\a\b\g\delta} \frac{\mathfrak{f}_{n,\a\b}^j \mathfrak{f}_{n,\g\delta}^j}{2} .
\end{equation}
where $\varepsilon^{\a\b\g\delta}$ is the Levi-Civita symbol and the sum runs over
repeated indices, and $\oint_{\boldsymbol{\theta}} \equiv \int_0^{2\pi} \! \frac{d^4 \boldsymbol{\theta}}{32 \pi^2}$. For $N=5$ terminal junctions, the topologies of the gapped Andreev bands are classified by the second Chern number Eq. \eqref{chern2}, which is analogous to the 4D quantum Hall effect. For $N=6$ terminal junctions we expect that SU(2) Yang monopoles form in the 5D Andreev bands by tuning the parameters of the $\hat{s}$-matrix. 

\section{Transport signals}\label{sec:transport} 

The current operator through the $\a$th lead is defined by $I_\a(\boldsymbol{\theta}) \equiv (2e/\hbar) \d_\a H(\boldsymbol{\theta})$. 
In the presence of constant voltages $\{V_\a\}$ applied to the leads, the SC phases vary linearly in time according to the second Josephson relation $\dot{\theta}_\a(t) =(2e/\hbar) V_\a$.     
The instantaneous eigenenergies and eigenstates are given by the ABS spectrum $\{E_n (\boldsymbol{\theta}(t))\}$ and wave functions $\{|\Phi_{n}(\boldsymbol{\theta}(t))\rangle\}$, respectively. 
We expand a wave function in the interaction representation
\begin{equation} \label{int-rep}
\left| \Psi(t) \right\rangle = \sum_{n,\k} e^{i \Theta_n(t)} c_n^\k (t)  | \Phi_n^\k(\boldsymbol{\theta}(t)) \rangle,
\end{equation}
where $\Theta_n(t) = -\frac{1}{\hbar} \int_{0}^{t}\! dt' E_{n}( \boldsymbol{\theta}(t'))$ is the dynamical phase, so that the Schr\"{o}dinger equation takes the form
\begin{align} 
\dot{c}_n^\k (t)= &\, - \sum_{\k'} \langle \Phi_n^\k | \dot{\Phi}_n^{\k'} \rangle c_n^{\k'} (t)  \nonumber \\
                  &\, - \frac{\hbar}{2e}\sum_{n' \neq n,\k'} c_{n'}^{\k'} (t) e^{i [\Theta^{n'}(t)-\Theta^n(t)]} \frac{\dot{\theta}_\a I_{\a,nn'}^{\k\k'}}{E_{n'}-E_n},    \label{sch-eq}
\end{align} 
where $I_{\a, nn'}^{\k\k'} \equiv \langle \Phi_n^\k |I_\a|\Phi_{n'}^{\k'} \rangle$ is the current matrix element in the instantaneous basis (see the Appendix \ref{sec:appendix} for further details). In the gapped phase, we impose the adiabatic condition 
$\mathrm{max}\{2eV_\a\} \ll \mathrm{min}\{ E_n-E_{n'} \}_{n \neq n'}$ and obtain
\begin{equation} \label{c-te}
\dot{c}_n(t) = -i \dot{\theta}_\a  \hat{\mathcal{A}}_{n,\a} \, c_n (t),
\end{equation}
where $c_n(t) \equiv (c_n^1(t), c_n^2(t))$ is a two-spinor in the degenerate space. The equation of motion Eq. \eqref{c-te} leads to $\d_\a c_n = -i \hat{\mathcal{A}}_{n,\a} \, c_n$, and, therefore, the non-Abelian nature of the Berry connection is manifested by $i[\partial_\alpha,\partial_\beta]c_n(t)=\hat{\mathcal{F}}_{\alpha\beta}c_n(t)$ in the interaction representation. The adiabatic time evolution gives $c_n(t) = \hat{U}_n(t) c_{n}(0)$, where  
\begin{equation} \label{def-U}
\hat{U}_n(t) \equiv \mathsf{P} e^{ -i \int_{\boldsymbol{\theta}(0)}^{\boldsymbol{\theta}(t)} \! d \theta_\a \, \hat{\mathcal{A}}_{n,\a}  (\boldsymbol{\theta})},   
\end{equation} 
with ``$\mathsf{P}$'' denoting the path order along the trajectory in $\boldsymbol{\theta}$ space. 

We assume an initial state that is the eigenstate of $H(\boldsymbol{\theta}(0))$ of energy $E_n(\boldsymbol{\theta}(0))$ as well as the adiabatic evolution [Eqs.~\eqref{int-rep} and \eqref{c-te}]. The instantaneous current through lead $\a$ reads $\mathcal{I}_{n,\a} (t) 
\equiv \langle \Psi^n(t) | I_\a(\boldsymbol{\theta}(t)) |\Psi^n(t) \rangle$, and by Eqs.~\eqref{int-rep} and \eqref{c-te} we obtain 
\begin{equation} \label{It}
\mathcal{I}_{n,\a} (t) = J_{n,\a} (\boldsymbol{\theta}(t)) + 2e  \big\langle \hat{\mathcal{F}}_{n,\a\b} \big\rangle \dot{\theta}_\b,
\end{equation}
where $J_{n,\a} (\boldsymbol{\theta}) \equiv \frac{2 e}{\hbar} \d_\a E_n(\boldsymbol{\theta})$ is the supercurrent and 
$\langle \cdots \rangle \equiv c_{n}^{\dagger}(t) \cdots  c_{n}(t)$ is the quantum mechanical average at time $t$. We note that the average Berry curvature contributes to the normal current and the instantaneous transconductance. Moreover, if we keep only $V_{2}$ finite and the other voltages vanishing, the time-averaged current, $\bar{\mathcal{I}}_1 = \mathcal{G}_{12} V_2,$ gives quantized transconductance, $\mathcal{G}_{12} = (4e^2/h) \mathcal{C}_{1}$,  that is proportional to the first Chern number $\mathcal{C}_{1}$ defined in the $\theta_{1,2}$ space, as extensively analyzed in earlier works \cite{Eriksson,Xie2,MH3}. 

In analogy to Eq.~\eqref{It}, the second Chern number is related to the time average of the instantaneous current correlation function  
$\mathcal{R}_{n,\a\b}(t) \equiv \langle \Psi^n(t) | \D I_\a^n \, \D I_\b^n |\Psi^n (t) \rangle$ with $\D I_\a^n(t) \equiv I_\a^n(t) - J_\a^n(t)$. We thus obtain
\begin{equation}\label{I2t}
\mathcal{R}_{n,\a\b}(t) = 4 e^2 \big\langle \hat{\mathcal{F}}_{n,\a\g} \hat{\mathcal{F}}_{n,\b\delta} \big\rangle \dot{\theta}_\g \dot{\theta}_\delta. 
\end{equation}
In the adiabatic limit, time averaging is equivalent to the integration through the entire phase space. We thus use the Josephson relation for dynamical phases and average autocorrelation function over $\boldsymbol{\theta}$. However, unlike in the case of a linear response, where already nonlocal conductance captures topological charge, a nonlinear response requires knowledge of all the nonlocal autocorrelations and only their properly symmetrized sum gives access to higher-rank topologies. Indeed, by combining Eqs. \eqref{chern2} and \eqref{I2t} we can extract 
\begin{equation}
\varepsilon^{\a\b\g\delta}\frac{\partial^2\bar{\mathcal{R}}_{\alpha\beta}}{\partial V_\gamma\partial V_\delta}=\left(\frac{4\pi e}{\phi_0}\right)^2\mathcal{C}_2,
\end{equation}
where the fundamental quantization unit is expressed via the flux quantum $\phi_0=h/2e$. This result tacitly assumes that in the expectation value for the average of the product of currents the cross-level terms give subleading contributions. Only then can the final result be expressed solely in terms of $\mathcal{C}_2$. At finite temperatures occupation functions of the bands would also enter the result.  We expect that the robustness of the quantization will be also limited by the voltage/phase noise. With the simplest assumption of white noise in the voltage sources leading to fluctuating phases, $\langle\delta\theta_\alpha(t)\delta\theta_{\beta}(t')\rangle=\Upsilon\delta_{\alpha\beta}\delta(t-t')$, described by a single broadening energy scale $\Upsilon$, one can estimate that the required measurement time to sufficiently average the current signals must exceed $(\Delta/eV)^2\Upsilon^{-1}$. Other limiting factors include Landau-Zener transitions between the bands and to the continuum of states above the gap leading to the dissipation. Finally, we note that phase dynamics can be described with the help of the Fokker-Planck equation that in particular yields the probability distribution function of phases. We leave a detailed analysis of these complications to the future work. 

\section*{Acknowledgments} 

We thank Matthew Foster for the communication regarding the symmetry classification of effective Hamiltonians and Vlad Pribiag for the discussions regarding Refs. \cite{Graziano,Pribiag}. The financial support for this work at the University of Wisconsin-Madison was provided by the National Science Foundation, U.S., Quantum Leap Challenge Institute for Hybrid Quantum Architectures and Networks, NSF Grant No. 2016136 (A. L.). The work of H.-Y.  X. was supported by the National Natural Science Foundation of China under Grant No. 12074039.  J. H. acknowledges support by the National Science Foundation, Grant No. DMR-1653661. 

\appendix 

\section{Adiabatic approximation} \label{sec:appendix}

In this section we sketch the derivation of the current correlation functions. The adiabatic approximation introduced in the Schr\"odinger equation \eqref{sch-eq} implies that $\mathcal{A}_{nn',\a}^{\k\k'}$ is dominated by the diagonal blocks $n=n'$, so that we introduce the formal decomposition $\hat{\mathcal{A}}_{nn',\a} = \delta_{nn'} \hat{\mathcal{A}}_{n,\a} + \l (1-\delta_{nn'}) \hat{a}_{nn',\a}$ where $\l \hat{a}_{nn',\a} = \hat{\mathcal{A}}_{nn',\a}$ and $\l \ll 1$. We define instantaneous eigenstates of the Hamiltonian 
\begin{equation} \label{ad-b}
| \eta_n^\k (t) \rangle \equiv e^{i \Theta_n(t)} \sum_{\k'} U_n^{\k'\k} (t)  | \Phi_n^{\k'}(\boldsymbol{\theta}(t)) \rangle,
\end{equation} 
where $\hat{U}(t)$ is the adiabatic evolution operator in Eq.~\eqref{def-U}. These satisfy the orthonormal condition $\langle \eta_{n'}^{\k'} (t)|\eta_n^\k (t) \rangle = \delta_{nn'}\delta_{\k\k'}$ and the completeness condition $1 = | \eta_n^\k (t) \rangle \langle \eta_n^\k (t)|$. For an arbitrary state in Eq.~\eqref{int-rep} we have 
\begin{equation} \label{proj}
\langle \eta_{n}^{\k}(t) | \Psi(t) \rangle = (\hat{U}_n^\dagger c_n)^\k. 
\end{equation}
Moreover, we obtain the important relations 
\begin{align}
&i\hbar|\dot{\eta}_n^\k \rangle = E_n |\eta_n^\k \rangle + i\hbar \dot{\theta}_\a \d_\a |\eta_n^\k \rangle, \nonumber \\
&i\d_\a |\eta_n^\k \rangle = e^{i \Theta_n(t)} \sum_{\k'} \big[(\hat{\mathcal{A}}_n \hat{U}_n)^{\k'\k} + U_n^{\k'\k} i\d_\a \big] | \Phi_n^{\k'} \rangle. \label{derivs}
\end{align}
In the basis \eqref{ad-b}, the matrix elements of the current operator reads
\begin{align}
\langle \eta_{n'}^{\k'} | I_\a |\eta_n^\k \rangle 
&\approx \frac{2e}{\hbar}\big[ \delta_{nn'}\delta_{\k\k'} \d_\a E_n \nonumber \\ 
&- i\hbar \langle \d_\a \eta_{n'}^{\k'}| \dot{\eta}_n^\k \rangle 
+ i\hbar \langle \dot{\eta}_{n'}^{\k'}|\d_\a \eta_n^\k \rangle     \big], \label{I-mat} 
\end{align}    
where we have used the adiabatic approximation of the form $H |\eta_n^\k \rangle  \approx i\hbar |\dot{\eta}_n^\k\rangle + \mathcal{O}(\l) $. Applying the relations in Eq.~\eqref{derivs} to Eq.~\eqref{I-mat}, we find
\begin{equation} \label{I-mat-f}
\langle \eta_{n'}^{\k'} | I_\a |\eta_n^\k \rangle 
\approx \delta_{nn'}  \left[\hat{U}_n^\dagger \left( J_\a^n + 2e \dot{\theta}_\b \hat{\mathcal{F}}_{n,\a\b} \right) \hat{U}_n \right]^{\k'\k},  
\end{equation} 
that is block diagonal up to small corrections in adiabaticity. Using the completeness condition with Eqs.~\eqref{proj} and \eqref{I-mat-f}, we obtain Eqs.~\eqref{It} and \eqref{I2t}. It is further possible to extend the calculation of the instantaneous current correlations to higher-order cumulants.  
  

%

\end{document}